# Prediction of superconductivity in Bilayer Kagome borophene

Yifan Han, Yue Shang, Wenhui Wan, Yong Liu, and Yanfeng Ge*

State Key Laboratory of Metastable Materials Science and Technology & Key Laboratory for Microstructural Material Physics of Hebei Province, School of Science, Yanshan University, Qinhuangdao, 066004, China

*Corresponding authors: yfge@ysu.edu.cn

The element boron has long been central to two-dimensional superconducting materials, and numerous studies have demonstrated the presence of superconductivity in various boron-based structures. Recent work introduced a new variant: Bilayer Kagome borophene, characterized by its bilayer Kagome lattice with van Hove singularity. Using first-principles calculations, our research investigates the unique electronic structure and superconducting properties of Bilayer Kagome borophene (BK-borophene) through first-principles calculations. BK-borophene is identified as a single-gap superconductor with an initial superconducting transition temperature ($T_c$) of 11.0 K. By strategically doping the material to align its Fermi level with the Van Hove singularity, $T_c$ is significantly enhanced to 30.0 K. The results contribute to the existing understanding of BK-borophene, highlighting its potential as a member of the expanding family of two-dimensional superconducting materials.

**INTRODUCTION**

The discovery of graphene has provided a groundbreaking platform for the application of nanomaterials, directing significant research attention toward the field of two-dimensional materials [1-2]. Among these materials, borophene is a notable member, anticipated to serve as a functional group or precursor for constructing boron nanotubes [3]. Boron (B) and nitrogen (N) have been pivotal components in numerous low-dimensional materials, attracting extensive interest from researchers over the past decade. Studies have revealed a rich diversity of two-dimensional compounds composed of carbon (C), boron (B), and nitrogen (N), including BN, $B_2C$, and $BC_2N$ [4]. However, previous research has indicated that only carbon forms a stable single-layer two-dimensional material, known as graphene. Nitrogen typically exists as a diatomic gas, while boron exhibits various three-dimensional structures [5]. The possibility of boron existing in a two-dimensional form has spurred both theoretical and experimental investigations into this elusive material. Theoretical studies suggest that a two-dimensional triangular lattice represents a metastable structure of borophene. A more stable form of borophene can be achieved by introducing vacancies into this triangular lattice, resulting in planar or curved geometries [7-11]. Borophene exhibits a high carrier density and a higher Young's modulus than graphene [12]. Additionally, its delocalized multicenter bonding confers metallic conductivity, distinct from the properties of bulk boron [13,14]. These remarkable mechanical properties and tunable anisotropic electronic characteristics position borophene as a promising candidate for future technological developments [5,15].

It is widely known that two-dimensional compounds play a crucial role in the development of nanosuperconducting quantum interference devices and nanosuperconducting transistors [16-21], provided they exhibit superconductivity. The induction of superconductivity in these non-metallic materials can be achieved by intercalating metal atoms. For example, graphene, silicene, and phosphorene are all semimetals or semiconductors with a zero density of states at the Fermi level, thereby lacking intrinsic superconductivity. The theoretical studies show superconducting transition temperatures (Tc) of 8.1 K for $LiC_6$ and 1.4 K for $CaC_6$ [22,23], and subsequent experimental results have confirmed transition temperatures of 5.9 K and 6.0 K for $LiC_6$ and $CaC_6$, respectively [24,25]. Similar investigations have been conducted on silicene and phosphorene, revealing superconducting transition temperatures of 15.5 K and 12.2 K for electron-doped silicene and phosphorene under specific tensile strains [26-28]. The prerequisite for inducing superconductivity in these materials involves doping to introduce charge carriers [26-28]. Conversely, research on the superconductivity of borophene has rapidly advanced, driven by its unique structural and metallic conductivity. Several polymorphs of borophene, including $\delta_6$, $\chi_3$, and $\beta_{12}$, have been identified, each demonstrating potential superconducting properties [29-32]. These findings suggest that borophene holds significant promise as a versatile material for developing next-generation superconductors, broadening the landscape of two-dimensional superconducting materials.

Recently, Qian Gao et al. proposed a stable two-dimensional crystal structure of Bilayer Kagome borophene (BK-borophene) composed of boron atoms [33], which also exhibits a Van Hove singularity (VHS) in its electronic structure. VHS phenomena are significant in condensed matter physics as they directly impact electronic transitions and play crucial roles in the electronic characteristics [34]. Typically, the appearance of VHS coincides with an enhanced electronic density of states. Recent studies on monolayer graphene and magic-angle graphene [35,36] have highlighted the role of VHS in facilitating superconductivity. BK-borophene displays metallic properties, similar to $\delta_6$, $\chi_3$, and $\beta_{12}$ borophene variants known for their superconducting behavior, and features VHS in its electronic structure. Therefore, investigating the potential for superconductivity in BK-borophene is highly pertinent. This work presents a comprehensive study of the structural stability and electronic properties of BK-borophene using first-principles calculations, with a specific focus on its superconducting properties and the influence of VHS.

**METHODS**

The first-principles calculations utilized Optimized Norm-Conserving Vanderbilt pseudopotentials [37] within the QUANTUM ESPRESSO package [38-39]. The Perdew–Burke–Ernzerhof parameterized generalized gradient approximation [40] was employed to describe the exchange correlation effect. A vacuum layer thickness of at least 20 Å along the c-axis was chosen to mitigate the crystal structure's periodicity effects and prevent interactions between adjacent atomic layers. The plane-wave cutoff energy was set to 50 Ry. Electronic structures and dynamical matrices were computed using a Monkhorst-Pack k-mesh of 24×24×1 and a q-grid of 12×12×1. During structure optimization, the force on each atom was constrained below $10^{-5}$ Ry/bohr, with the total energy fluctuation kept within $10^{-6}$ Ry to ensure stability. To assess the crystal structure's stability, phonon spectral calculations and molecular dynamics simulations were conducted. Maximally localized Wannier functions [41-44] required for EPW calculations were interpolated using an unshifted Brillouin-zone k-mesh of 24×24×1. The anisotropic Migdal-Eliashberg equations were solved with interpolated k- and q-point grids of 120×120×1, sufficiently dense to converge the superconducting gaps. Electrons and phonons were treated with smearing functions replacing Dirac δ-functions, using widths of 15 meV and 0.2 meV, respectively. The superconducting transition temperature was evaluated using the McMillian-Allen-Dynes formula [45-48].

# RESULTS AND DISCUSSION

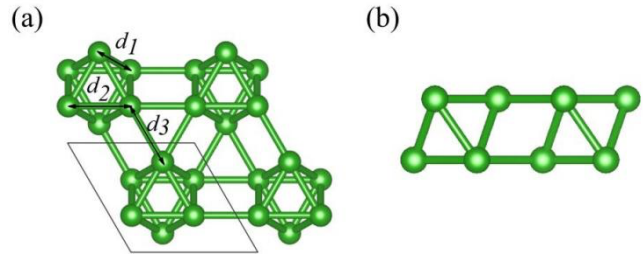

FIG. 1. (a) Top and (b) side views of BK-borophene. The black lines indicate a primitive unit cell. The green spheres denote boron atoms. Three B-B bond lengths are present in the structure, denoted as $d_1$, $d_2$, and $d_3$ respectively.

The Kagome lattice is a two-dimensional hexagonal arrangement composed of three interlinked triangular sublattices sharing common edges. The previous works has proposed a stable structure of a new boron form called Bilayer Kagome borophene (BK-borophene) by substituting the sublattice [33]. The smallest member of the closed borane family, $[B_6H_6]^{2-}$, has been previously prepared and extensively studied, showing that BK-borophene can be produced through polymerization [49]. Figure 1 illustrates the top and side views of BK-borophene (space group: P3m1), with lattice constants of a = b = 3.55 Å. The BK-borophene lattice comprises three distinct B-B bond lengths, labeled as $d_1$, $d_2$, and $d_3$, measuring 1.77 Å, 1.73 Å, and 1.82 Å, respectively. Each atom is six-coordinated, forming completely planar upper and lower layers, separated by a distance of 1.46 Å. As depicted in Figure 2(a), the phonon dispersion for BK-borophene shows no imaginary frequencies across the entire Brillouin zone, indicating dynamic stability. Additionally, molecular dynamics simulations conducted at room temperature using a 4×4×1 supercell to minimize the effects of periodic boundary conditions demonstrate that the total energy remains stable, oscillating within a reasonable range without significant variations [Figure 2(b)]. This confirms the thermodynamic stability of the BK-borophene.

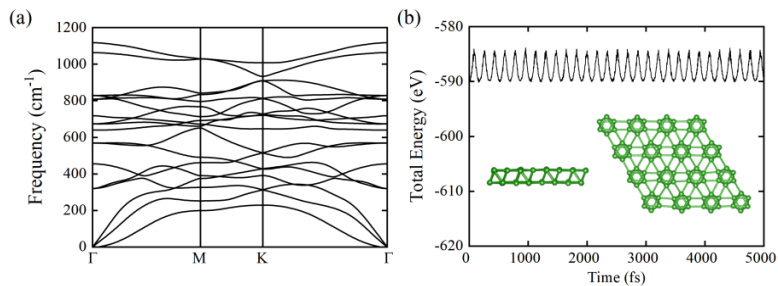

FIG. 2. (a) Phonon spectra of BK-borophene. (b) AIMD simulation of BK-borophene at 300 K. The simulation processes 5000 steps at 1 fs per step.

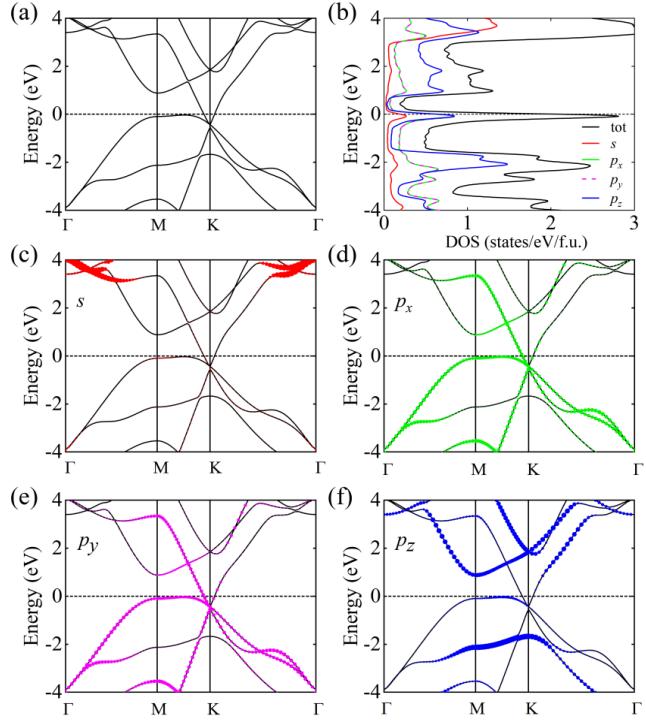

FIG. 3. (a) The electronic band structures and (b) projective electronic density of states of BK-borophene. And projected electronic band structures for (c) s,(d) px,(e) py and (f) pz orbitals.

Figure 3 presents the electronic band structure of BK-borophene, clearly indicating its metallic nature due to the presence of electronic states at the Fermi level. To further investigate the electronic properties of BK-borophene, we analyzed the projected density of states (PDOS) for the s, $p_x$, $p_y$, and $p_z$ orbitals. As shown in Figure 3(b), the results are consistent with the electronic band structure, confirming the metallic properties of BK-borophene. Additionally, the Fermi level is found to be close to VHS, which results in a significant increase in the DOS as well as positively impacts the superconductivity. Figures 3(c-f) illustrate the relative contributions of different orbitals to the DOS at the Fermi level. It is evident that the B-s orbital contributes minimally to the state density at the Fermi level, while the B-p orbitals contribute substantially. Among the p orbitals, the px, py, and pz orbitals contribute almost equally to the Fermi level, highlighting their significant role in the electronic properties of BK-borophene.

The superconductivity of BK-borophene is investigated due to its metallic properties and the proximity of its Fermi level to VHS, both of which favor superconductivity. Figure 4 displays the calculated phonon dispersion, phonon density of states (PHDOS), and Eliashberg spectral function $\alpha^2F(\omega)$ for BK-borophene. The phonon frequencies can be categorized into high and low-frequency regions. Notably, the frequency-dependent coupling $\lambda(\omega)$ reveals that atomic vibrations in the low-frequency range of 0-400 cm$^{-1}$ significantly contribute to the electron-phonon coupling (EPC), accounting for approximately 55% of the electron-phonon coupling constant λ (λ = 0.737). Figure 4(a) illustrates the phonon dispersion with phonon linewidth γ$_{qv}$ represented by red bubbles, where the bubble size correlates with the magnitude of γ$_{qv}$. The optical phonons exhibit relatively flat dispersion, resulting in notable peaks in PHDOS and λ(ω) at frequencies of 320 cm$^{-1}$, 718 cm$^{-1}$, and 1064 cm$^{-1}$, therefore it is evident that these phonons contribute substantially to the EPC evidently. Figure 4(c) depicts the atomic vibration modes at these frequencies, with the arrow sizes and directions indicating the amplitude and direction of atomic vibrations, respectively. At 320 cm$^{-1}$, atoms vibrate oppositely along the B-B bond direction, resembling vibrations along a circular cutting edge. At 718 cm$^{-1}$, atoms exhibit nearly opposite vibrations along out-of-plane direction. Furthermore, at 1064 cm$^{-1}$, atoms in one layer vibrate inward while those in the other layer vibrate outward, causing a diffusive motion in the lower layer when the upper layer contracts. These detailed phonon and atomic vibration analyses provide insights into the significant electron-phonon interactions that underlie the superconducting properties of BK-borophene. The superconducting transition temperature Tc is defined as the temperature at which the superconducting gap vanishes. As shown in Figure 4(b), the superconducting gap decreases with increasing temperature and disappears at 11.0 K, indicating a Tc of 11.0 K for BK-borophene. At T = 5 K, the superconducting energy gap of BK-borophene ranges from 1.41 meV to 1.75 meV, demonstrating that it is a single-gap superconducting material.

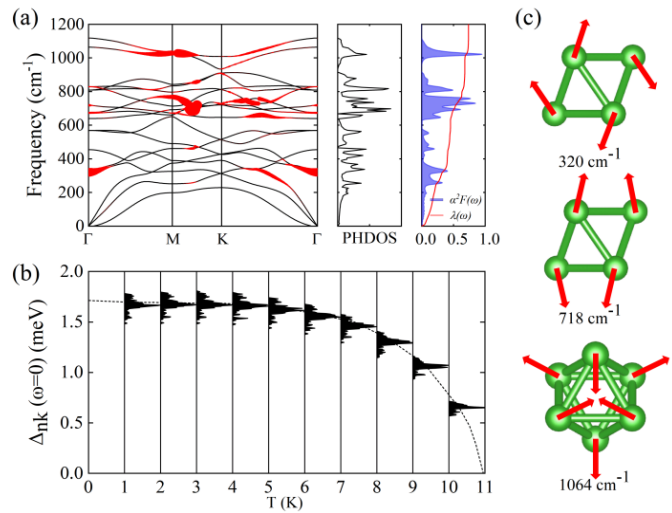

FIG. 4. (a) Phonon dispersion with phonon linewidth γ$_{qv}$ in red bubble, phonon density of states (PHDOS), isotropic Eliashberg spectral functions α$^2$F(ω) and cumulative frequency-dependent coupling λ(ω). (b)The variation of the superconducting gap (Δ$_{nk}$) with temperature is obtained by solving the anisotropic Eliashberg equation with μ∗= 0.1. (c) The vibration patterns of the phonon modes, and the length of the arrows indicates the amplitude of atomic motions.

The DOS for BK-borophene (Figure 3) shows VHS at 0.08 eV below the Fermi level. Previous theoretical studies have also highlighted the benefit of VHS in enhancing superconductivity. As shown in Figure 5(a), doping with the concentration of 0.1 holes/cell shifts the Fermi level to VHS, increasing DOS at the Fermi level from 2.13 states/eV/f.u. to 2.74 states/eV/f.u. Moreover, Figure 5(b) demonstrates that there are no imaginary frequencies in the phonon dispersion curve of BK-borophene after hole doping, indicating that the stability is maintained under hole doping. To further investigate the effect of VHS on superconductivity, λ and Tc are recalculated using the McMillan-Allen-Dynes equation. As shown in Figure 6(a), when the Fermi level intersects the VHS, the value of α²F(ω) increases in the 0~234 cm$^{-1}$ range, with a pronounced peak at 212 cm$^{-1}$, which leads to an increase in λ from 0.737 to 0.899. Figure 6(b) shows that while the number of superconducting gaps in BK-borophene remains unchanged, the magnitude of the superconducting gap increases. For instance, at 10 K, the superconducting gap increases by approximately 10 meV. Ultimately, the superconducting gap vanishes at 30.0 K, raising the Tc of BK-borophene from 11.0 K to 30.0 K under the influence of hole doping at a concentration of 0.1 holes/cell.

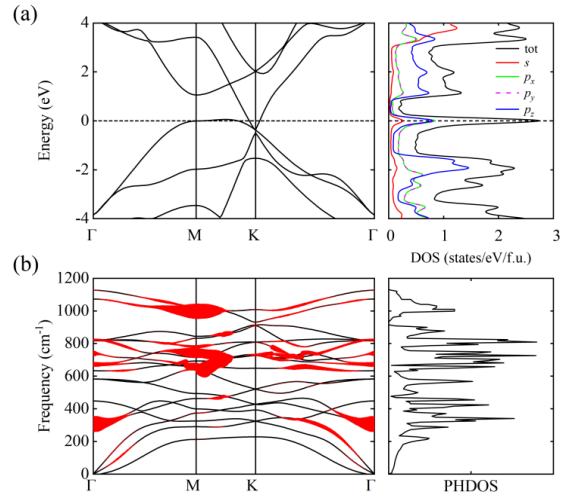

FIG. 5. (a) The electronic band structures and projective electron-band density of states, (b) Phonon dispersion with phonon linewidth γqv in red bubble and phonon density of states (PHDOS) of BK-borophene after hole-doping.

**CONCLUSION**

Our work presents a comprehensive study of the electronic structure and superconductivity of BK-borophene using first-principles calculations. The results indicate that BK-borophene is a single-gap superconducting material with a superconducting transition temperature of 11.0 K. Notably, when the Fermi level is adjusted to pass through the Van Hove singularity via hole doping, Tc increases significantly to 30.0 K. This study not only enhances our understanding of BK-borophene's properties but also expands the family of two-dimensional superconducting materials. Furthermore, it demonstrates that the superconducting transition temperature of BK-borophene can be effectively increased by leveraging the Van Hove singularity through hole doping.

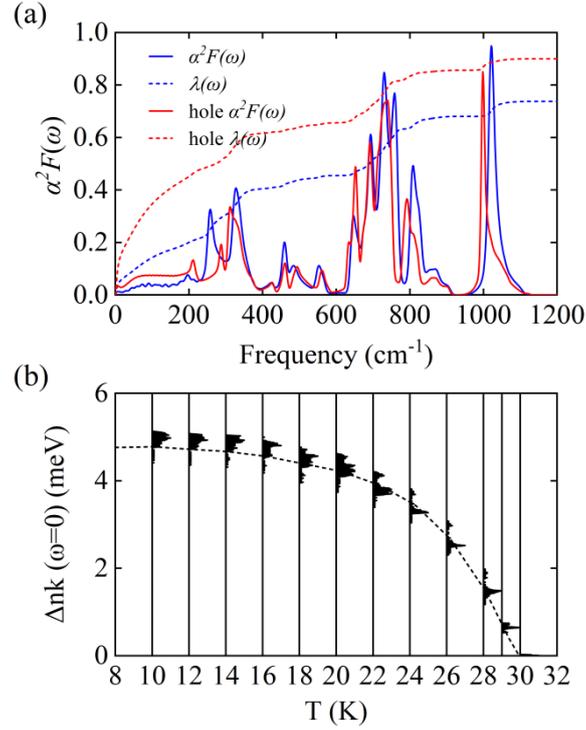

FIG. 6. (a) Isotropic Eliashberg spectral functions α2F(ω) and cumulative frequency-dependent coupling λ(ω). (b) The variation of the superconducting gap ($\Delta$nk) with temperature of BK-borophene after hole-doping is obtained by solving the anisotropic Eliashberg equation with $\mu* = 0.1$.

## ACKNOWLEDGMENTS

This work is supported by the Innovation Capability Improvement Project of Hebei province (22567605H), the Natural Science Foundation of Hebei Province (A2022203006), the Science and Technology Project of Hebei Education Department (BJK2022002).